\documentstyle{article}
\newlength{\mytopmargin}
\newlength{\myleftmargin}
\begin{document}
\title{Some exact  correlations in the
 Dyson Brownian motion model for transitions to the CUE }

\author{P.J. Forrester\thanks{email: matpjf@maths.mu.oz.au; supported by the ARC
}\\
Department of
Mathematics\\ University of Melbourne\\ Parkville\\ Victoria 3052\\ Australia}
\date{}
\maketitle
\begin{abstract} The Dyson Brownian motion model for transitions to the CUE
is considered. For initial eigenvalue probability density functions
corresponding to the COE and CSE, the density-density correlation function
between an eigenvalue at position $\lambda'$ initially ($\tau = 0$)
and an eigenvalue at position $\lambda$ for  general $\tau$, is calculated. 
 Theoretical predictions for the
asymptotic behaviour are verified, and  the distribution for the initial
rate of change of the eigenvalues is calculated. Also, for initial conditions corresponding to
equi-spaced eigenvalues, the $n$-point equal parameter distribution is
computed. The extension of the calculation of these quantities to a
multiparameter version of the theory is given.
\end{abstract}
\section{Introduction}
A problem of fundamental importance in studying quantum aspects of classically
chaotic systems is the statistical properties of the energy spectra. Since the
pioneering work of Wigner and Dyson (see e.g. [1]) it has been realized that
the statistical properties are determined by the presence, or not, of a 
global time reversal symmetry $T$. Details of the matrix elements of the 
Hamiltonian are unimportant, apart from the global constraint
 imposed by the symmetry $T$. Thus the discrete portion of the energy spectrum
can be studied by modelling the Hamiltonian by a random Hermitian matrix if
there is no time reversal symmetry, and in the case of a time reversal
symmetry by a random real symmetric matrix if $T^2 = 1$ and a random real
quaternion matrix if $T^2 = -1$.

More complicated situations can arise. In particular the quantum system may 
contain a parameter which can continuously vary the spectrum within a
universality class or indeed between universality classes (see e.g.~[2,3]). 
For the latter
situation, by considering parameter ($\tau$) dependent Gaussian random
matrices (see e.g.~[4]), it is possible to show that the eigenvalue
probability density function (p.d.f.) $p(\lambda_1, \dots, \lambda_N;\tau)$
satisfies the Fokker-Planck equation
$$
\gamma {\partial p \over \partial \tau} ={\cal L} p, \qquad {\cal L} =
\sum_{j=1}^N {\partial  \over \partial \lambda_j}
\left ( {\partial W \over \partial \lambda_j}
+\beta^{-1} {\partial  \over \partial \lambda_j} \right )
\eqno (1.1)
$$
where
$$
W = 
- \sum_{1 \le j < k \le N} \log |\lambda_j - \lambda_k| + {1 \over 2}
\sum_{j=1}^N \lambda_j^2.
\eqno (1.2) 
$$
Here   $\gamma$ provides a scale for
$\tau/\lambda_j^2$ and $\beta = 1,2$ or 4 depending on whether the transition is to
orthogonal, unitary or symplectic symmetry. With $\tau = 0$, $p$ is specified
as the eigenvalue p.d.f.~of one of the three Gaussian random matrix ensembles,
with a symmetry different to that after the transition. In the random matrix
context, the Fokker-Planck equation (1.1) is referred to as the 
Dyson Brownian motion model.

The objective of this paper is to provide the exact evaluation of some
two-point correlation functions for the transition to unitary symmetry.
In Section 2 we revise exact results on this problem which are already
known [5,6], as well as a theoretical prediction for asymptotic behaviours, 
the implication this prediction has for the integrated correlator of the
variance of a linear statistic [7], and the calculation of the
distribution function for the initial rate of change
of the eigenvalues. In Section 3 we provide the exact
evaluation for the density-density correlation between an eigenvalue at
$\tau = 0$ and an eigenvalue at general $\tau$, when the eigenvalue p.d.f~at
$\tau = 0$ corresponds to orthogonal symmetry, while in Section 4 the
calculation will be repeated with the $\tau = 0$ eigenvalue p.d.f.~corresponding
to symplectic symmetry. In both cases the expected asymptotic behaviour is
verified, and the distribution function for the initial rate of
change of the eigenvalues is calculated. In Section 5 we calculate the equal-parameter $n$-point
correlations when the initial conditions correspond to equispaced eigenvalues,
while in Section 6 we show how the results of the previous sections can be
extended to apply to the multiple parameter case.

\section{Summary of known results}
\subsection{Equal parameter distributions}
The so called equal-parameter distribution functions
$$
\rho_n(\lambda_1, \dots, \lambda_N;\tau) :=
N(N-1) \dots (N - n) \left ( \prod_{l = n+1}^N \int_{-\infty}^\infty
d \lambda_l \right ) p(\lambda_1, \dots, \lambda_N;\tau)
\eqno (2.1)
$$
have been calculated exactly for the Fokker-Planck equation (1.1) with
$\beta = 2$, and initial conditions
$$
 p(\lambda_1, \dots, \lambda_N;0) = {1 \over C_{1N}}e^{-W(\lambda_1,
\dots,\lambda_N)}
\eqno (2.2a)
$$
$$
 p(\lambda_1, \dots, \lambda_N;0) = {1 \over C_{4(N/2)}}
e^{-4W(\lambda_2,\lambda_4,
\dots,\lambda_{N})} \prod_{l = 1}^{N/2}\delta(\lambda_{2l} - \lambda_{2l - 1})
\qquad (N \:{\rm even})
\eqno (2.2b)
$$
in refs.~[5,6].

We are particularly interested in these distribution functions in the
thermodynamic limit, when $N \rightarrow \infty$ while the density of
eigenvalues is fixed at some non-zero constant $\rho$. In the
$N \rightarrow \infty$ limit, the
density implied by (1.1) is (see e.g.~[4]) the Wigner semi-circle
$$
\rho(\lambda) \: \sim \: {2 \sqrt{2N}\over \pi } \sqrt {1 - \lambda^2 / 2N},
$$
so to obtain a bulk density $\rho$ about the origin it is necessary to scale
the eigenvalues $\lambda_j \mapsto \pi \rho \lambda_j/ \sqrt {2 N }$. This can
be achieved by instead scaling  $\gamma$ so that
$$\gamma \mapsto
2 N\gamma / (\pi \rho)^2 .
\eqno (2.3)
$$
 With this scale, the thermodynamic limit of the
equal time distributions (2.1) exist and can be computed exactly for the initial
conditions (2.2). In particular, for the initial conditions (2.2), the two-point
distribution function is given by [5,6] (for consistency with our
notation  the parameter $\rho^2$ in [5] is
identified with $\rho^2 \tau/2 \gamma$ and $r$ replaced by $\rho \lambda$)
\renewcommand{\theequation}{2.4} 
\begin{eqnarray}
\rho_{(2)}(\lambda_1,\lambda_2) & = & \rho^2 - {\rho^2 \over 2} {\rm Tr}
\left [ \begin{array}{cc}a(\lambda) & a_{12}^{\beta_0}(\lambda, \tau) \\
 a_{21}^{\beta_0}(\lambda, \tau) & a(\lambda) \end{array} \right ]
\left [ \begin{array}{cc}a(\lambda) & -a_{12}^{\beta_0}(\lambda, \tau) \\
 -a_{21}^{\beta_0}(\lambda, \tau) & a(\lambda) \end{array} \right ] \nonumber \\
& = & \rho^2 - \rho^2\Big ( a(\lambda)^2 - a_{12}^{\beta_0}(\lambda, \tau)
a_{21}^{\beta_0}(\lambda, \tau) \Big )
\end{eqnarray}
where
$$
\lambda := \lambda_1 - \lambda_2 \qquad a(\lambda) := {\sin \pi \rho \lambda
\over \pi \rho \lambda }
$$
and for the initial condition (2.2a) ($\beta_0 = 1$)
$$
 a_{12}^{1}(\lambda, \tau) := - \pi \int_0^1  du\, u \sin \pi u \rho \lambda
\, e^{\pi^2 \rho^2 \tau u^2 /  \gamma}
\eqno (2.5a)
$$
$$
 a_{21}^{1}(\lambda, \tau) := -{1 \over \pi} \int_1^\infty du \,{\sin \pi u \rho \lambda
\over u} e^{-\pi^2 \rho^2 \tau u^2 /  \gamma}, 
\eqno (2.5b)
$$
while for the initial condition (2.2b)
$$
 a_{12}^{4}(\lambda, \tau) := - \pi \int_1^\infty du \,u \sin \pi u \rho \lambda
\, e^{-\pi^2 \rho^2 \tau u^2 /  \gamma}
\eqno (2.6a) 
$$
$$
 a_{21}^{4}(\lambda, \tau) := -{1 \over \pi} \int_0^1  du \, {\sin \pi u \rho \lambda
\over u} \, e^{\pi^2 \rho^2 \tau u^2 /  \gamma}.
\eqno (2.6b)
$$
\subsection{ Density-density and current-current correlations}
To define the density-density correlation, $S(\lambda, \lambda';\tau)$ say,
between an eigenvalue $\lambda'$ at $\tau = 0$ and an eigenvalue $\lambda$
at general $\tau$, let
$$
G(\lambda_1', \dots, \lambda_N'|\lambda_1, \dots, \lambda_N;\tau)
$$
denote the Green's function for the Fokker-Planck equation (1.1) i.e.~the
solution which satisfies the initial condition
$$
G(\lambda_1', \dots, \lambda_N'|\lambda_1, \dots, \lambda_N;0)
= \prod_{l = 1}^N \delta (\lambda_l - \lambda_l'),
\eqno (2.7)
$$
and let $p_0(\lambda_1', \dots, \lambda_N')$ denote the initial eigenvalue p.d.f.,
and $\rho_{(1)}(\lambda,\tau)$ denote the one-point distribution (the
density) defined by
(2.1). Then
\renewcommand{\theequation}{2.8} 
\begin{eqnarray}
S(\lambda, \lambda';\tau) & := & \Big ( \prod_{l = 1}^N \int_{-\infty}^\infty 
d \lambda'_l \Big ) \sum_{j=1}^N \delta(\lambda_j' - \lambda') \,
p_0(\lambda_1', \dots, \lambda_N') \nonumber \\
& & \times \Big ( \prod_{l = 1}^N \int_{-\infty}^\infty 
d \lambda_l \Big ) \sum_{j=1}^N \delta(\lambda_j - \lambda) \,
G(\lambda_1', \dots, \lambda_N'|\lambda_1, \dots, \lambda_N;\tau)
- \rho_{(1)}(\lambda',0)\rho_{(1)}(\lambda,\tau). \nonumber \\
\end{eqnarray}
In the thermodynamic limit, with the scaling (2.3),
$$
\rho_{(1)}(\lambda,\tau) = \rho
$$
for all $\tau$ and we expect $S$ to depend only on the difference $\lambda
-\lambda'$, so we write
$$
S(\lambda, \lambda';\tau) := S(\lambda - \lambda';\tau).
$$

On the basis of a non-local diffusion equation approximation to (1.1)
Beenakker and Rejaei [7] have obtained a prediction for the small-$k$
behaviour of the Fourier transform
$$
\hat{S} (k;\tau) := \int_{-\infty}^\infty S(\lambda - \lambda';\tau)
e^{i k (\lambda - \lambda')} d \lambda.
$$
This prediction states that
$$
\hat{S} (k;\tau) \: \sim \: \hat{S} (k;0) e^{-\pi^2 \rho^2 \tau |k| /  \gamma}
\qquad {\rm as } \quad {k \rightarrow 0}.
\eqno (2.9)
$$
In particular, if the initial p.d.f.~is 
$$
{1 \over C_{\beta N}} e^{-\beta W}
\eqno (2.10)
$$
where $W$ is given by (1.2), then [1]
$$
\hat{S} (k;0) \: \sim \: {|k| \over \pi \beta} \qquad {\rm as} \quad |k|
\rightarrow 0,
$$
so taking the inverse transform of (2.9) gives
$$
S(\lambda - \lambda';\tau) \: \sim \:
{1 \over 2 \pi^2 \beta} {\partial^2 \over \partial \lambda^2}
\log \Big ( (\pi\rho \tau / \gamma)^2 + (\lambda - \lambda')^2 \Big )
\qquad {\rm for} \quad
 |\lambda - \lambda'| \rightarrow \infty.
\eqno (2.11)
$$ 

When the initial eigenvalue p.d.f.~is (2.10) with $\beta = 2$ (the Fokker-Planck
equation then describes a perturbation of the unitary ensemble), the
density-density correlation is well known (see e.g.~[7]):
$$
S(\lambda - \lambda';\tau) =  \rho^2\int_0^1 du_1 \, \cos \pi u_1 \rho \lambda \,
e^{\pi^2 \rho^2 \tau u_1^2 / 2 \gamma} 
\int_1^\infty du_2 \cos \pi u_2 \rho \lambda \,
e^{-\pi^2 \rho^2 \tau u_2^2 / 2 \gamma} .
\eqno (2.12)
$$
(with the eigenvalues $\lambda$ and $\lambda'$ chosen in the neighbourhood of the
edge of the Wigner semi-circle, the exact evaluation of $S$ is also known [8]).
The asymptotic expansion of (2.12) agrees with (2.11).

In this paper we will provide the exact evaluation of $S(\lambda - \lambda';\tau)$
for the initial conditions  (2.2). 
For the  initial conditions (2.2a) the above theory gives
$$
\hat{S} (k;\tau) \: \sim \: {|k| \over \pi }
 e^{-\pi^2 \rho^2 |k| \tau /  \gamma}
 \qquad {\rm as} \quad |k|
\rightarrow 0,
\eqno (2.13a)
$$
while for the  initial conditon (2.2b)
$$
\hat{S} (k;0) \: \sim \: 2{|k| \over 4 \pi  } \qquad {\rm as} \quad |k|
\rightarrow 0,
$$
(the factor of 2 accounts for the double degeneracy in (2.2b)), and so (2.9)
gives
$$ 
\hat{S} (k;\tau) \: \sim \: {|k| \over 2\pi }
 e^{-\pi^2 \rho^2 |k| \tau /  \gamma}
 \qquad {\rm as} \quad |k|
\rightarrow 0.
\eqno (2.13b)
$$

Now consider the correlator $R(\tau;\alpha)$ of a linear statistic
$$
A(\tau; \alpha) := \sum_{j=1}^N a(\lambda_j(\tau) / \alpha).
$$
The correlator is defined as
\renewcommand{\theequation}{2.14}
\begin{eqnarray}
R(\tau;\alpha) & := & \Big \langle(A(0;\alpha) - \langle A(0;\alpha) \rangle)
(A(\tau;\alpha) - \langle A(\tau;\alpha) \rangle) \Big \rangle \nonumber \\
& = & \int_{-\infty}^\infty d \lambda \int_{-\infty}^\infty d \lambda'
a(\lambda / \alpha) a(\lambda'/\alpha)S(\lambda - \lambda';\tau),
\end{eqnarray}
where the averages in the first line above are over both the eigenvalues at
$\tau = 0$ and the eigenvalues at general $\tau$. Beenakker and
Rejaei  [7] used the asymptotic formula (2.9) to compute the integrated
correlator
$$
\chi_A := \int_0^\infty dx R(x^2;\alpha),
$$
for the initial p.d.f.~(2.10),
in the limit $\alpha \rightarrow \infty$ (the quantity $x^2 = \tau$ is identified
in [7] as an external perturbation parameter). Thus, introducing Fourier
transforms, we have
$$
\chi_A = {\alpha \over 2 \pi} \int_0^\infty dx \int_{-\infty}^\infty
dk \, |\hat{a}(k)|^2\hat{S} (k/\alpha;x^2).
$$
In the limit $\alpha \rightarrow \infty$, use of (2.9) gives, after carrying out 
the integration over $x$, (eq.~(4.22) of [7] with $\gamma \mapsto
\gamma / \pi \rho_0$)
$$
\chi_A = {1 \over 2 \pi^2 \beta }\left ( { \gamma \over \pi \rho^2} \right )^{1/2}
\int_0^\infty dk \, |\hat{a}(k)|^2 k^{1/2}.
$$

Another consequence of (2.9) is a formula for the current-current
correlation, defined formally as
$$
C(\lambda, \lambda';\tau) := \Big \langle
\sum_{j,k =1}^N {d \lambda_j(0) \over dx} {d \lambda_k(\tau) \over dx}
\delta (\lambda - \lambda_j(0)) \delta (\lambda - \lambda_k(\tau)) \Big \rangle
$$
 In a Brownian motion model the derivatives cannot be
computed directly. However, by considering the correlator (2.14) with
$A$ replaced by $dA / dx$, it can be shown  that the current-current
correlation is related to the density-density correlation by [7, eq. 4.14]
$$
\hat{C} (k; x^2) = -{1 \over k^2}{\partial^2 \over \partial x^2}
\hat{S}(k;x^2).
\eqno (2.15)
$$
Thus, from (2.9), for small-$|k|$ and the initial condition (2.10), one obtains
[7]
$$
\hat{C} (k; x^2) \: \sim \: {2 \rho \over \beta \gamma} (1 - 2 \pi^2
\rho^2 |k| x^2/\gamma)e^{-\pi^2
\rho^2 |k| x^2/\gamma},
$$
or equivalently
$$
C(\lambda, \lambda';x^2) :=  C(\lambda - \lambda';x^2)\: \sim \:
{1 \over 2 \pi^2 \beta}
{\partial^2 \over \partial x^2 } \log \Big ( (\pi \rho x^2 /\gamma)^2
+ (\lambda - \lambda')^2 \Big )
\eqno (2.16)
$$
for large-$|\lambda - \lambda'|$.

\subsection{The distribution of the initial rate of change of the 
eigenvalues}

As commented in the above section, the parameter
$ x := \tau^{1/2} $ can be identified with an external perturbing parameter.
For the initial conditions (2.2b) a quantity of interest [9] associated 
with this interpretation is the distribution of the 
scaled repulsion between degenerate levels as the
perturbation is initialized:
\renewcommand{\theequation}{2.17}
\begin{eqnarray} A(\alpha) & := & \lim_{\tau \rightarrow 0} \lim_{N \rightarrow
\infty} \Big \langle {2 \over N} \sum_{j=1}^{N/2}
\delta \Big ( \alpha - {\lambda_{2j}(\tau) - \lambda_{2j-1}(\tau) \over x}
\Big ) \Big \rangle \nonumber \\  & = &
\lim_{\tau \rightarrow 0}\lim_{N \rightarrow
\infty} \Big \langle {1 \over 2 \pi} {2 \over N} \sum_{j=1}^{N/2}
\int_{-\infty}^\infty dp \, e^{ip(\alpha - (\lambda_{2j}(\tau) - \lambda_{2j-1}
(\tau))/\tau^{1/2})} \Big \rangle \nonumber \\
& = & \lim_{\tau \rightarrow 0}{1 \over 2 \pi \rho} 
\int_{-\infty}^\infty dp \, e^{ip\alpha}\int_{-\infty}^\infty d\lambda \,
\rho^T_{(2)}(\lambda; \tau)e^{-ip \lambda/\tau^{1/2}}\nonumber \\
& = & \lim_{\tau \rightarrow 0}{ \tau^{1/2} \over \rho}
\rho^T_{(2)}(\alpha \tau^{1/2}; \tau)
\end{eqnarray} 
The exact result (2.4) gives [9]
$$
 A(\alpha) = \Big ( {\gamma \over 2 \pi} \Big )^{1/2} \gamma \alpha^2
e^{- \gamma \alpha^2/2}.
\eqno (2.18)
$$

A quantity of interest for general initial conditions is the distribution of
the initial rate of change of the eigenvalues with respect to $x$:
\renewcommand{\theequation}{2.19}
\begin{eqnarray}
f(v) & := &  \lim_{\tau \rightarrow 0}
\Big \langle {1 \over N} \sum_{j=1}^N \delta ( v - {\lambda_j (\tau) - \lambda_j(0)
\over x}  ) \Big \rangle \nonumber \\
& = & \lim_{\tau \rightarrow 0} {1 \over 2 \pi \rho}
\int_{-\infty}^\infty dp \, e^{ip v} \int_{-\infty}^\infty d\lambda \,
S(\lambda, \tau) e^{- i \lambda p /\tau^{1/2}} \nonumber \\
& = & \lim_{\tau \rightarrow 0} { \tau^{1/2} \over \rho}
S(v\tau^{1/2}; \tau)
\end{eqnarray}
Using the exact result (2.4) for initial conditions with unitary symmetry,
(2.19) gives [10] the Gaussian distribution
$$
f(v) = \Big ( {\gamma \over 2 \pi} \Big )^{1/2} e^{- v^2 \gamma / 2}.
\eqno (2.20)
$$

\section{Exact solution for initial conditions
with orthogonal symmetry }
\subsection{Green's function for $\beta = 2$ in periodic model}
Our interest is in calculating the correlation (2.8) in the thermodynamic limit.
For this purpose we can replace the potential $W$ in (1.2) by its periodic
version
$$
W = - \sum_{1 \le j < k \le N} \log |\sin \pi (\lambda_k - \lambda_j)/L|, \qquad
{- L/2 \le \lambda_j \le L/2}.
\eqno (3.1)
$$
This gives the Fokker-Planck equation of the Dyson Brownian motion model for the
circular ensemble of unitary random matrices [11].

In general the equilibrium solution of (1.1) is
$$
p = {1 \over C_{N \beta}} e^{- \beta W},
\eqno (3.2)
$$
and the operator
$$
e^{\beta W / 2} {\cal L} e^{- \beta W / 2}
$$
is Hermitian [12]. With $\beta = 2$ and $W$ given by (3.1), a standard calculation
gives
$$
-e^{\beta W / 2} {\cal L} e^{- \beta W / 2} = -{1 \over 2}
\sum_{j = 1}^N {\partial^2 \over \partial \lambda_j^2} - E_0
\eqno (3.3)
$$
where
$$
E_0 = \left ({2 \pi \over L} \right )^2 {N(N^2 - 1) \over 12}.
$$
The well known reason for the tractibility of the case $\beta = 2$ is thus
revealed: the Hermitian 
operator on the r.h.s.~of (3.3) is the Schr\"odinger operator
for non-interacting particles. To reclaim the correct equilibrium solution
(3.2) the particles are assumed to be  impenetrable, so that the
wavefunction vanishes if two particles are at the same point.

Thus if $g(\lambda_1^{(0)}, \dots, \lambda_N^{(0)}| \lambda_1, \dots,\lambda_N;
\tau)$ denotes the Green's function satisfying the imaginary time  Schr\"odinger
equation 
$$
-{\partial g \over \partial \tau} = -{1 \over 2} \sum_{j = 1}^N 
{\partial^2 \over \partial \lambda_j^2} - E_0,
\eqno (3.4)
$$
then the Green's function $G$ for the Fokker-Planck equation (1.1), with $\beta = 2$
and $W$ given by (3.1), is related to $g$ by
\renewcommand{\theequation}{3.5}
\begin{equation}
G(\lambda_1^{(0)}, \dots, \lambda_N^{(0)}| \lambda_1, \dots,\lambda_N;
\tau)
 = e^{-(W( \lambda_1,
 \dots,\lambda_N)-W(\lambda_1^{(0)}, \dots, \lambda_N^{(0)}))}
g(\lambda_1^{(0)}, \dots, \lambda_N^{(0)}| \lambda_1, \dots,\lambda_N;
\tau)
\end{equation}
Due to the impenetrability condition of the last paragraph we can suppose that
the eigenvalues are ordered
$$
-L/2 \le \lambda_1 < \dots < \lambda_N \le L/2,
\eqno (3.6)
$$
(and similarly the initial eigenvalues $\lambda_j^{(0)}$), which
 allows us to write
$$
e^{-(W( \lambda_1, \dots,\lambda_N)-W(\lambda_1^{(0)}, \dots, \lambda_N^{(0)}))}
= \prod_{1 \le j < k \le N} { \sin \pi (\lambda_k - \lambda_j)/L
\over \sin \pi (\lambda_k^{(0)} - \lambda_j^{(0)})/L}.
\eqno (3.7)
$$
For $N$ even (3.7) is anti-periodic under the mapping $\lambda_k \mapsto \lambda_k
+ L$. We then require that $g$ be antiperiodic under
the same mapping, so by (3.5), $G$ will be periodic.

It is straightforward to check [6,13] that for $N$ even the Green's function solution
of (3.4) satisfying the required conditions is
\renewcommand{\theequation}{3.8}
\begin{equation}
g(\lambda_1^{(0)}, \dots, \lambda_N^{(0)}| \lambda_1, \dots,\lambda_N;
\tau)
 = e^{ \tau E_0 / \gamma} 
\det[{1 \over L} \theta_2(\pi(\lambda_j - \lambda_k^{(0)})/L;q)]_{j,k = 1,\dots,N}
\end{equation}
where
$$
q := e^{- 2 \pi^2 \tau / \gamma L^2} \qquad {\rm} \qquad
\theta_2(z;q) := \sum_{n = -\infty}^\infty q^{(n - 1/2)^2}e^{2 i z (n-1/2)},
\eqno (3.9)
$$
(for $N$ odd $\theta_2$ is to be replaced by
$\theta_3$).
Thus we have that for $N$ even the required Green's function  is given by
substituting (3.7) and (3.8) in (3.5).

\subsection{The integration technique}
We want to calculate the density-density correlation (2.8) with $G$ given as
above and the initial condition
$$
p(\lambda_1^{(0)}, \dots, \lambda_N^{(0)}) = {1 \over C_{1N}}
 \prod_{1 \le j < k \le N}  |\sin \pi (\lambda_k^{(0)} - \lambda_j^{(0)})/L|
\eqno (3.10) 
$$
Since we are considering the periodic model the integrations in (2.8)
must now be from $-L/2$ to $L/2$ and 
$$
\rho_{(1)}(\lambda';0) = \rho_{(1)}(\lambda;\tau)=N/L = \rho.
$$
The exact value of the normalization $ C_{1N}$ is known but will not be needed in
our subsequent analysis.

To calculate (2.8) we will consider the generalized partition function
$$
Z(a,b) := \prod_{l=1}^N\left ( \int_{-L/2}^{L/2} d\lambda^{(0)}_l\,
b(\lambda^{(0)}_l) \int_{-L/2}^{L/2}d\lambda_l\, a(\lambda_l) \right )
p(\lambda_1^{(0)}, \dots, \lambda_N^{(0)})
G(\lambda_1^{(0)}, \dots, \lambda_N^{(0)}| \lambda_1, \dots,\lambda_N;
\tau)
\eqno (3.11)
$$
and use the formula
$$
S(\lambda, \lambda';\tau) =
{\delta^2 \over \delta a(\lambda) \delta b(\lambda')}
\log Z(a,b) \Big |_{a = b =1}.
\eqno (3.12)
$$
Substituting for $p$ and $G$ we have
$$
Z(a,b) =
C \prod_{l=1}^N\left ( \int_{-L/2}^{L/2} d\lambda^{(0)}_l\,
b(\lambda^{(0)}_l) \int_{-L/2}^{L/2}d\lambda_l\, a(\lambda_l) \right )
 \prod_{1 \le j < k \le N} {\rm sgn} (\lambda^{(0)}_k - \lambda^{(0)}_j)
\hspace{1.5cm}
$$
$$
\times \det[{1 \over L} \theta_2(\pi(\lambda_j
 - \lambda_k^{(0)})/L;q)]_{j,k = 1,\dots,N} \det[e^{2 \pi i \lambda_j (k - (N +1)/2)/L}]_{j,k = 1,\dots,N},
\eqno (3.13)
$$
where $C$ is independent of $\lambda_l^{(0)}, \lambda_l$, and the 
Vandermonde determinant formula
$$
 \prod_{1 \le j < k \le N} 2 \sin \pi (\lambda_k - \lambda_j)/L
= i^{- N(N-1)/2}\det[e^{2 \pi i \lambda_j (k - (N +1)/2)/L}]_{j,k = 1,\dots,N}
\eqno (3.14)
$$
has been used.

To simplify (3.13), we note that since
$$
\det[{1 \over L} \theta_2(\pi(\lambda_j
 - \lambda_k^{(0)})/L;q)]_{j,k = 1,\dots,N}
$$
is anti-symmetric in each $\lambda_l$, we can make the replacement
$$
\det[e^{2 \pi i \lambda_j (k - (N +1)/2)/L}]_{j,k = 1,\dots,N}
\mapsto N! \prod_{j=1}^N e^{2 \pi i \lambda_j (j - (N +1)/2)/L}
$$
in the integrand. The integrations over the $\lambda_l$ can now be performed 
row-by-row to give
$$
Z(a,b) =
C N!\prod_{l=1}^N\left ( \int_{-L/2}^{L/2} d\lambda^{(0)}_l\,
b(\lambda^{(0)}_l) \right )
 \prod_{1 \le j < k \le N} {\rm sgn} (\lambda^{(0)}_k - \lambda^{(0)}_j)
\det[A_j(\lambda_k^{(0)})]_{j,k = 1,\dots,N}
\eqno (3.15)
$$
where
$$
A_j(\lambda_k^{(0)}) :=
{1 \over L} \int_{-L/2}^{L/2} d\lambda\, a(\lambda) \theta_2(\pi(\lambda
 - \lambda_k^{(0)})/L;q) e^{2 \pi i \lambda (j - (N +1)/2)/L}.
\eqno (3.16)
$$ 
The integrations over the $\lambda^{(0)}_l$ can now be done using the method
of integration over alternate variables [1] to give
$$
Z(a,b) =
C N!^2\det[H_{jk}]_{j,k = 1,\dots,N}^{1/2}
\eqno (3.17)
$$
where
$$
H_{jk} := \int_{-L/2}^{L/2} d\lambda^{(0)}\, b(\lambda^{(0)})
\,\int_{-L/2}^{L/2} d\lambda^{(0)}_1\,b(\lambda^{(0)}_1)
A_j(\lambda^{(0)})A_k(\lambda_1^{(0)})
{\rm sgn} (\lambda^{(0)} - \lambda^{(0)}_1).
\eqno (3.18)
$$

Since
$$
\rho = {1 \over 2 \det[H_{jk}]} {\delta \over \delta a(\lambda)} \det[H_{jk}]
\Big |_{a = b =1} = {1 \over 2 \det[H_{jk}]} {\delta \over \delta b(\lambda')}
 \det[H_{jk}]
\Big |_{a = b =1}
\eqno (3.19)
$$
(3.12) and (3.17) give
$$
S(\lambda, \lambda';\tau) = -2\rho^2 + {1 \over 2 \det[H_{jk}]}
{\delta^2 \over \delta a(\lambda) \delta b(\lambda')}\det[H_{jk}]
\Big |_{a = b =1}.
\eqno (3.20)
$$
\subsection{Calculation of the determinants}
From (3.16), and the definition of $\theta_2$, when $a=b=1$
$$
A_j(\lambda) = e^{2 \pi i (j - (N+1)/2)\lambda / L}
q^{(j-(N+1)/2)^2}.
\eqno (3.21)
$$
With this value of $A_j(\lambda)$ we have
$$
\int_{-L/2}^{L/2} d \lambda \, A_j(\lambda){\rm sgn} (\lambda - \lambda')
=-{L \over \pi i (j - (N+1)/2 )} A_j(\lambda')
\eqno (3.22)
$$
Substituting (3.21) and (3.22) in (3.18) shows that when $a=b=1$
$$
H_{jk} = h_j
\delta_{N+1-j-k,0},
\eqno (3.23a)
$$
where
$$
h_j := -{L^2 \over \pi i (j - (N+1)/2)}q^{2(j-(N+1)/2)^2}
\eqno (3.23b)
$$
Hence $[H_{jk}]_{j,k = 1, \dots,N}$ is only non-zero along the right-to-left
 diagonal.

Taking the functional derivative $\delta / \delta b(\lambda')
$ in row $j$ and setting $a=b=1$  gives
\renewcommand{\theequation}{3.24}
\begin{eqnarray}
I_{j,k}(\lambda') & := & \int_{-L/2}^{L/2} d \lambda \,
\Big ( -A_j(\lambda)A_k(\lambda') + A_j(\lambda')A_k(\lambda) \Big ){\rm sgn}
(\lambda' - \lambda) \nonumber \\
& = & e^{2 \pi i (j + k - (N + 1))\lambda'/L} q^{(j-(N+1)/2)^2 + (k - (N + 1)/2)^2}
{L \over \pi i } \left ( {1 \over k - (N + 1)/2} - {1 \over j - (N +1)/2}
\right ). \nonumber \\
\end{eqnarray}
Taking the functional derivative $\delta / \delta a(\lambda)
$ in row $j'$ and setting $a=b=1$  gives, for $j' \ne j$
\renewcommand{\theequation}{3.25}
\begin{eqnarray}
J_{j',k}(\lambda) & := & {1 \over L}\int_{-L/2}^{L/2} d \lambda^{(0)} 
\int_{-L/2}^{L/2} d \lambda^{(0)}_1 \,
\Big ( \theta_2(\pi(\lambda - \lambda^{(0)})/L;q)
e^{2 \pi i (j' - (N +1)/2) \lambda /L}A_k(\lambda^{(0)}_1) \nonumber \\
& & + \theta_2(\pi(\lambda - \lambda^{(0)}_1)/L;q)
e^{2 \pi i (k - (N +1)/2) \lambda /L}A_j(\lambda^{(0)}) \Big )
{\rm sgn}(\lambda^{(0)} - \lambda^{(0)}_1) \nonumber \\
& = &  q^{2(k-(N+1)/2)^2} e^{2 \pi i (j' + k - (N + 1))\lambda/L}
\left ( { L \over \pi i (k - (N + 1)/2)} \right ) - (j' \leftrightarrow k),
\end{eqnarray}
while for $j' = j$ we obtain
\renewcommand{\theequation}{3.26}
\begin{eqnarray}\lefteqn{
K_{j,k}(\lambda, \lambda')}\nonumber \\ & := & {1 \over L}
  \theta_2(\pi(\lambda - \lambda')/L;q)\int_{-L/2}^{L/2} d \lambda^{(0)}
\Big (e^{2 \pi i (j - (N +1)/2) \lambda /L}A_k(\lambda^{(0)}) 
- (j \leftrightarrow k) \Big ){\rm sgn}(\lambda' - \lambda^{(0)}) \nonumber \\
& & +\Big (e^{2 \pi i (j - (N +1)/2) \lambda /L}A_k(\lambda') 
- (j \leftrightarrow k) \Big ){1 \over L}\int_{-L/2}^{L/2} d \lambda^{(0)}
{\rm sgn}( \lambda^{(0)} - \lambda')
  \theta_2(\pi(\lambda - \lambda^{(0)})/L;q) \nonumber \\
& = & {1 \over L}
  \theta_2(\pi(\lambda - \lambda')/L;q)\Big ({ L \over \pi i (k - (N + 1)/2)}
q^{(k-(N+1)/2)^2} e^{2 \pi i (j\lambda + k\lambda' - (\lambda + \lambda')(N + 1)/2)/L}
- (j \leftrightarrow k) \Big )\nonumber \\
&  & + \Big (q^{(k-(N+1)/2)^2}
 e^{2 \pi i (j\lambda + k\lambda' - (\lambda + \lambda')(N + 1)/2)/L}
- (j \leftrightarrow k) \Big )
{1 \over L}\int_{-L/2}^{L/2} d \lambda^{(0)}
{\rm sgn}( \lambda^{(0)} - \lambda')
  \theta_2(\pi(\lambda - \lambda^{(0)})/L;q) \nonumber \\  
\end{eqnarray}

From (3.23), with $a = b =1$, in each row $j''$ $(\ne j,j')$ the
 only non-zero term is along the right to left diagonal and is given by $h_{j''}$.
Expanding the determinant along this diagonal shows that (3.20) reduces to
$$
S(\lambda',\lambda;\tau) = -2 \rho^2 +{1 \over 2} \bigg (
\sum_{j,j' =1 \atop j \ne j'}^N
\Big ({I_{j,N+1-j}(\lambda')J_{j',N+1-j'}(\lambda) \over h_j h_{j'}}
-  (j \leftrightarrow j', N+1-j(j')\mbox{ fixed}) \Big )
$$
$$
+ \sum_{j=1}^N {K_{j,N+1-j}(\lambda,\lambda') \over h_j} \bigg ).
\eqno (3.27)
$$
We note that the above expression is unchanged if we remove the restriction
$j \ne j'$ in the double sum. Doing this, and noting
$$
 \sum_{j=1}^N {I_{j,N+1-j}(\lambda') \over h_j}
= \sum_{j'=1}^N {J_{j',N+1-j'}(\lambda) \over h_j'} = 2 \rho,
$$
(3.27) reads
$$
S(\lambda',\lambda;\tau) = {1 \over 2} \bigg (
 \sum_{j=1}^N {K_{j,N+1-j}(\lambda,\lambda') \over h_j}-
\sum_{j,j' =1}^N 
{I_{j,N+1-j'}(\lambda')J_{j',N+1-j}(\lambda) \over h_j h_{j'}}
 \bigg ).
\eqno (3.28)
$$

The quantities in the above expressions are given explicitly by (3.23b) and 
(3.24)-(3.26). We see immediately that the double sum is a Riemann approximation
to a definite integral, with the approximation becoming exact in the
thermodynamic limit. Taking the thermodynamic limit of the first term requires
an intermediate step: from (3.26) the limiting behaviour of $\theta_2$
is required. 

From the Poisson summation formula
$$
\sum_{n = -\infty}^\infty e^{- \epsilon (n - 1/2)^2} e^{2 i (n - 1/2) z}
= \left ( { \pi \over \epsilon} \right )^{1 / 2} \sum_{m = -\infty}^\infty (-1)^m
e^{-(z + \pi m)^2 / \epsilon}.
$$
Hence with $z = \pi (\lambda - \lambda')/L$ and
 $\epsilon = 2 \pi^2 \tau / \gamma L^2$
\renewcommand{\theequation}{3.29}
\begin{eqnarray}
{1 \over L}\theta_2(\pi(\lambda - \lambda')/L;q) 
& = &{1 \over L} \left ( \gamma L^2 \over 2 \pi \tau \right )^{1/2}
 \sum_{m = -\infty}^\infty (-1)^m
e^{-\gamma  (\lambda - \lambda' +  mL)^2 / 2\tau} \nonumber \\
& \sim &  \left ( \gamma \over 2 \pi \tau \right )^{1/2}
e^{-\gamma  (\lambda - \lambda' )^2 / 2\tau} \qquad {\rm as} \quad L
\rightarrow \infty \nonumber \\
& = & \rho \int_{- \infty}^\infty du_1 \, e^{- 2 \tau (\pi \rho u_1)^2/\gamma}
\cos 2 \pi u_1 \rho (\lambda - \lambda').
\end{eqnarray}

Using this asymptotic formula, and from the remarks in the paragraph above,
we thus have that in the thermodyamic limit
\renewcommand{\theequation}{3.30}
\begin{eqnarray}\lefteqn{
S(\lambda',\lambda;\tau)=} \nonumber \\ &  &
\rho^2 \int_{- \infty}^\infty du_1 \, e^{- 2 \tau (\pi \rho u_1)^2/\gamma}
\cos 2 \pi u_1 \rho (\lambda - \lambda')
 \int_{- 1/2}^{1/2} du_2 \, e^{ 2 \tau (\pi \rho u_1)^2/\gamma}
\cos 2 \pi u_2 \rho (\lambda - \lambda') \nonumber \\
& & + \rho^2 \int_{- \infty}^\infty du_1 \,{ e^{- 2 \tau (\pi \rho u_1)^2/\gamma}
\over u_1}
\sin 2 \pi u_1 \rho (\lambda - \lambda')
 \int_{- 1/2}^{1/2} du_2 \,u_2 e^{ 2 \tau (\pi \rho u_1)^2/\gamma}
\sin 2 \pi u_2 \rho (\lambda - \lambda') \nonumber \\
& & -{\rho^2 \over 2}  \int_{- 1/2}^{1/2} du_1  \int_{- 1/2}^{1/2} du_2
(u_1 + u_2) \Big ( {1 \over u_1} e^{- 2 \tau (\pi \rho)^2 (u_1^2-u_2^2)/\gamma}
+ {1 \over u_2} e^{- 2 \tau (\pi \rho)^2 (u_2^2-u_1^2)/\gamma} \Big )
e^{2 \pi i (u_1 - u_2) \rho (\lambda' - \lambda)} \nonumber \\
\end{eqnarray}
where to obtain the second term the formula
$$
 \int_{- \infty}^\infty d\lambda^{(0)} {\rm sgn}(\lambda^{(0)}
- \lambda')\int_{- \infty}^\infty du_1 \, e^{- 2 \tau (\pi \rho u_1)^2/\gamma}
\cos 2 \pi u_1 \rho (\lambda^{(0)} - \lambda)
$$
$$
= - {1 \over \pi}\int_{- \infty}^\infty du_1 \,{ e^{- 2 \tau (\pi \rho u_1)^2/\gamma}
\over u_1}
\sin 2 \pi u_1 \rho (\lambda - \lambda')
$$
has been used.

Expressing the double integrals as single integrals and simplifying gives
\renewcommand{\theequation}{3.31}
\begin{eqnarray}
\lefteqn{S(\lambda',\lambda;\tau) : =S(\lambda'-\lambda;\tau)} \nonumber \\
& = & \rho^2 \Big (\int_1^\infty du_1 \, e^{-  \tau (\pi \rho u_1)^2/2\gamma}
\cos  \pi u_1 \rho (\lambda - \lambda')
\int_0^1 du_2 \, e^{  \tau (\pi \rho u_2)^2/2\gamma}
\cos  \pi u_2 \rho (\lambda - \lambda') \nonumber \\
& + & \int_{1}^\infty du_1 \,{ e^{-  \tau (\pi \rho u_1)^2/2\gamma}
\over u_1}
\sin  \pi u_1 \rho (\lambda - \lambda')
 \int_{0}^{1} du_2 \,u_2 e^{  \tau (\pi \rho u_2)^2/2\gamma}
\sin  \pi u_2 \rho (\lambda - \lambda')
\end{eqnarray}
We note that the first term is precisely the density-density correlation 
for initial conditions with unitary symmetry given by (2.11), while the
second term is precisely that which occurs in the two-point
distribution function (2.4) (after replacing $\tau$ by $2\tau$) 
with $\beta_0 = 1$ (after replacing $\tau$ by $2 \tau$).
The large $\tau$, $\lambda - \lambda'$ expansion of the above integrals is obtained
by expanding the integrands in the neighbourhood of $u_1,u_2 = 1$. The
behaviour (2.1) with $\beta = 1$ is verified.

The Fourier transform with respect to $\lambda - \lambda'$ of each of the
products in (3.31) have been calculated in ref.~[7] and [5] respectively.
Thus we have
$$
\hat{S}(k;\tau) = \hat{S}_1(k;\tau) + \hat{S}_2(k;\tau)
\eqno (3.32a)
$$
where
$$
 \hat{S}_1(k;\tau) := {\gamma \over \tau \pi |k|}
\exp (- \tau \pi |k| q_{{\rm max}} \rho /\gamma)
{\rm sinh} (\tau \pi |k| q_{{\rm min}} \rho /\gamma)
\eqno (3.32b)
$$
with
$$
q_{{\rm min}} : = {\rm min} (1,|k|/2 \pi \rho)
 \qquad q_{{\rm max}} := {\rm max}(1,|k|/2 \pi \rho),
$$
and
$$
\hat{S}_2(k;\tau) := {\gamma \over 2 \pi \tau |k|}(1 - e^{-\tau k^2/\gamma})
-{|k| \over 2 \pi} e^{\tau k^2/2 \gamma} \int_1^{1 +|k|/\pi \rho}
dk_1 \, {1 \over k_1} e^{- \pi \tau \rho |k| k_1/\gamma},
\qquad 0 \le |k| \le 2 \pi \rho,
\eqno (3.32c)
$$
$$
\hat{S}_2(k;\tau) := {\gamma \over  \pi \tau |k|} e^{-\tau k^2/2 \gamma}
\sinh \pi \tau \rho |k|/\gamma
-{|k| \over 2 \pi} e^{\tau k^2/2 \gamma} \int_{-1 + |k|/ \pi \rho}^{1 +|k|/\pi \rho}
dk_1 \, {1 \over k_1} e^{- \pi \tau \rho |k| k_1/\gamma},
\qquad |k| \ge 2 \pi \rho.
\eqno (3.32d)
$$
Although the small-$|k|$ behaviour 
of this expression agrees with (2.13a), we see that there is
also a singularity at $|k| = 2\pi \rho$. This corresponds to
oscillatory terms in the large-$|\lambda - \lambda'|$ expansion of (3.31)
and indicates a preference for crystalline ordering with spacing $1/2 \pi \rho$.
 
For $\tau, |\lambda - \lambda'| \rightarrow 0$ (3.31) gives
$$
S(\lambda'-\lambda;\tau) \: \sim \: {\rho \over \tau^{1/2}}
\int_0^\infty du_1 \, e^{- u_1^2 /
 2\gamma} \cos \pi u_1 (\lambda - \lambda')/\tau^{1/2}
$$
Thus, from (2.17),
\renewcommand{\theequation}{3.33}
\begin{eqnarray}
f(v) & = & \int_0^\infty du_1 \, e^{- u_1^2 / 2\gamma} \cos \pi u_1 v
\nonumber \\
& = &  \Big ( {\gamma \over 2 \pi} \Big )^{1/2} e^{- v^2 \gamma / 2},
\end{eqnarray}
which is identical to the result (2.20) for initial conditions with
unitary symmetry. It is also identical to the result found by Simons and
Altshuler [10] when the initial and final distributions are given by (2.2a)
(i.e.~for a perturbation of the eigenvalue p.d.f.~with orthogonal
symmetry).
\section{Exact solution with symplectic symmetry initial conditions}
The Green's function for the Dyson Brownian motion model for the circular
ensemble with $\beta = 2$ and $N$ even is given by substituting
 (3.7) and (3.8) in (3.5), and
is of course valid independent of the particular initial conditions. In this
section we want
to calculate the density-density correlation with the initial conditions
(2.2b).

The first step is to take the limits
$$
\lambda_{2j-1}^{(0)} \mapsto \lambda_{2j}^{(0)} \qquad j=1, \dots,N/2
$$
in the Green's function. We readily find, assuming the ordering (3.6)
\renewcommand{\theequation}{4.1}
\begin{eqnarray}
\lefteqn{G = \left ( {L \over \pi} \right )^N e^{\tau E_0 /\gamma}
{\prod_{1 \le j < k \le N} \sin \pi (\lambda_k - \lambda_j)/L
\over\prod_{1 \le j < k \le N/2} \sin^4 \pi (\lambda_{2k}^{(0)} - 
\lambda_{2j}^{(0)})/L}}
\nonumber \\
& & \times \det [\begin{array}{cc}
{1 \over L}\theta_2(\pi(\lambda_j - \lambda^{(0)}_{2k})/L;q) &
{ {\partial \over \partial \lambda_{2k}^{(0)}}}
{1 \over L}\theta_2(\pi(\lambda_j - \lambda^{(0)}_{2k})/L;q)\end{array}]_{j=1,
\dots,N \atop k = 1, \dots,N/2}
\end{eqnarray}
To obtain an explicit expression for the generalized partition function
(3.11) we multiply $G$ by
$$
p(\lambda_2^{(0)}, \dots, \lambda_N^{(0)}) = {1 \over C_{4(N/2)}}
 \prod_{1 \le j < k \le N/2}  \sin^4 \pi 
(\lambda_{2k}^{(0)} - \lambda_{2j}^{(0)})/L. 
$$
This gives
\renewcommand{\theequation}{4.2}
\begin{eqnarray}
\lefteqn{
Z(a,b) := \prod_{l=1}^N\left ( \int_{-L/2}^{L/2} d\lambda^{(0)}_l\,
b(\lambda^{(0)}) \int_{-L/2}^{L/2}d\lambda_l\, a(\lambda_l) \right )}
 \nonumber \\
& & \times  \det [\begin{array}{cc}
{1 \over L}\theta_2(\pi(\lambda_j - \lambda^{(0)}_{2k})/L;q) &
{ {\partial \over \partial \lambda_{2k}^{(0)}}}
{1 \over L}\theta_2(\pi(\lambda_j - \lambda^{(0)}_{2k})/L;q)\end{array}]_{j=1,
\dots,N \atop k = 1, \dots,N/2} \det[e^{2 \pi i \lambda_j
 (k - (N +1)/2)/L)}]_{j,k = 1,\dots,N} \nonumber \\
\end{eqnarray}
where we have used the Vandermonde determinant formula (3.14).

The integrations over $\lambda_1, \dots, \lambda_N$ can be performed
row-by-row while the integrations over
 $\lambda_2^{(0)}, \dots, \lambda_N^{(0)}$ yield a pfaffian structure [1].
 We obtain
$$
Z(a,b) = C \left ( \det [ \int_{-L/2}^{L/2} d\lambda^{(0)} b(\lambda^{(0)})
U_{jk}(\lambda^{(0)})]_{j,k = 1,\dots,N} \right )^{1/2},
\eqno (4.3)
$$
where
$$
U_{jk}(\lambda^{(0)}) :=
A_j(\lambda^{(0)}) {\partial \over \partial \lambda^{(0)} }A_k (\lambda^{(0)})
- (j \leftrightarrow k)
$$
and $A_j$ is given by (3.16). 
The density-density correlation can be calculated from (4.3) using 
(3.20) with $H_{jk}$ replaced by $U_{jk}$ and $-2\rho^2$
replaced by $-\rho^2$. In fact the calculation is very
similar to that presented in Section 3.3. 

Thus with $a=b=1$
$$
U_{jk} =u_j \delta_{N+1-j - k, 0}
\eqno (4.4a)
$$
where
$$
u_j : = 2 \pi i (k-j) q^{2 (j-(N + 1)/2)^2},
\eqno (4.4b)
$$
so that $[U_{jk}]_{j,k = 1, \dots,N}$ is non-zero only along the right-to-left
diagonal (c.f. (3.23)).
Taking the functional derivative $ {\delta / \delta a(\lambda)}
$ in row $j$ and setting $a=b=1$  gives
$$
V_{jk}(\lambda) := {4 \pi i \over L} e^{2 \pi i \lambda (j + k - (N+1))/L}
\Big ( (k - (N+1)/2))q^{2(k - (N+1)/2)^2} - (j \leftrightarrow k) \Big )
\eqno (4.5)
$$
while taking the functional derivative $ {\delta / \delta b(\lambda')}
$ in row $j'$ gives
$$
W_{j'k}(\lambda') = {2 \pi i \over L}(k-j')q^{(k - (N+1)/2  )^2+(j' - (N+1)/2  )^2} 
e^{2 \pi i \lambda' (j' + k - (N+1))/L}
\eqno (4.6)
$$
for $j \ne j'$, and
\renewcommand{\theequation}{4.7}
\begin{eqnarray}\lefteqn{ X_{jk}(\lambda,\lambda') } \nonumber \\
&=& {1 \over L} e^{2 \pi i \lambda (j  - (N+1)/2)/L} 
\theta_2(\pi(\lambda - \lambda')/L;q){\partial \over
 \partial \lambda' }A_k (\lambda')
- (j \leftrightarrow  k) \nonumber \\
& & +{1 \over L} e^{2 \pi i \lambda (k  - (N+1)/2)/L}
 e^{2 \pi i \lambda' (j  - (N+1)/2)/L}q^{(j - (N+1)/2  )^2}
{\partial \over \partial \lambda' }
\theta_2(\pi(\lambda - \lambda')/L;q) - (j \leftrightarrow k)
\end{eqnarray}
for $j = j'$.

Since in each row $j''$ $(\ne j,j')$ the only non-zero element in the
denominator of (3.20) (with $H_{jk}$ replaced by $U_{jk}$) is along the 
right-left diagonal and is given by $u_{j'}$, we conclude (c.f~(3.28))
$$
S(\lambda',\lambda;\tau) = {1 \over 2} \bigg (
 \sum_{j=1}^N {X_{j,N+1-j}(\lambda,\lambda') \over u_j}-
\sum_{j,j' =1}^N
{V_{j,N+1-j'}(\lambda)W_{j',N+1-j}(\lambda') \over u_j u_{j'}}
-  (j \leftrightarrow j') \bigg )
\eqno (4.8)
$$
where we have used the fact that
$$
 \sum_{j=1}^N {V_{j,N+1-j}(\lambda) \over u_j}
= 2  \sum_{j'=1}^N {W_{j',N+1-j'}(\lambda') \over u_j'} = 2 \rho
$$
The thermodynamic limit of (4.8) is computed  analogously to that of (3.28).
We find
\renewcommand{\theequation}{4.9}
\begin{eqnarray}
\lefteqn{S(\lambda',\lambda;\tau) : =S(\lambda'-\lambda;\tau)} \nonumber \\
& = &{1 \over 2} \rho^2 \Big (\int_1^\infty du_1 \, e^{-  \tau(\pi \rho u_1)^2/2\gamma}
\cos  \pi u_1 \rho (\lambda - \lambda')
\int_0^1 du_2 \, e^{  \tau (\pi \rho u_2)^2/2\gamma}
\cos  \pi u_2 \rho (\lambda - \lambda') \nonumber \\
& + & \int_{1}^\infty du_1 \, u_1 e^{-  \tau (\pi \rho u_1)^2/2\gamma}
\sin  \pi u_1 \rho (\lambda - \lambda')
 \int_{0}^{1} du_2 \, { e^{  \tau (\pi \rho u_2)^2/2\gamma} \over u_2 }
\sin  \pi u_2 \rho (\lambda - \lambda')
\end{eqnarray}
This expression is very similar to that for the orthogonal symmetry initial
condition given by (3.30). In particular, 
apart from the factor of $1/2$, the first term is the  density-density
correlation (2.12), while the second term occurs in the expression (2.4)
with $\beta_0 = 4$ for the two-point distribution. Also,
the method of obtaining the 
large-$|\lambda - \lambda'|$, $\tau$ expansion is the same
as in (3.30). Behaviour corresponding to (2.13b) is
verified (here it is essential that the large-$\tau$ limit also be included, for
otherwise the leading-$|\lambda - \lambda'|$ expansion is oscillatory
and decays $O(1/|\lambda - \lambda'|)$,
analogous to the $\tau = 0$ result [1]).

The Fourier transform with respect to $\lambda - \lambda'$ of (4.9) is readily
calculated. We find
$$
\hat{S}(k;\tau) = {1 \over 2}\Big (\hat{S}_1(k;\tau) + \hat{S}_3(k;\tau) \Big )
\eqno (4.10a)
$$
where $\hat{S}_1(k;\tau)$ is given by (3.32b) and
$$
\hat{S}_3(k;\tau) := {\gamma \over  \pi \tau |k|}
 e^{- \pi \tau \rho |k| /\gamma}
\sinh \pi \tau \rho |k|/\gamma
+{|k| \over 2 \pi} e^{-\tau k^2/2 \gamma} \int^1_{1 -|k|/\pi \rho}
dk_1 \, {1 \over k_1} e^{- \pi \tau \rho |k| k_1/\gamma},
\qquad 0 \le |k| \le 2 \pi \rho,
\eqno (4.10b)
$$
$$
\hat{S}_3(k;\tau) := {\gamma \over  \pi \tau |k|} e^{-\tau k^2/2 \gamma}
\sinh \pi \tau \rho |k|/\gamma
+{|k| \over 2 \pi} e^{-\tau k^2/2 \gamma} \int_{-1 }^{1}
dk_1 \, {1 \over k_1} e^{- \pi \tau \rho |k| k_1/\gamma},
\qquad |k| \ge 2 \pi \rho.
\eqno (4.10c)
$$
Thus $\hat{S}(k;\tau)$ has a logarithmic singularity at $|k| = 2\pi \rho$.
The first singularity corresponds to the leading oscillatory term
of the large-$|\lambda - \lambda'|$ expansion of (4.9) which has period
$1 /\pi \rho$ (the initial spacing between the doubly degenerate
pairs of eigenvalues), while the second singularity corresponds to
the preferred crystalline ordering of spacing $1/2 \pi \rho$ between
eigenvalues in the final state.
 
For $\tau, |\lambda - \lambda'| \rightarrow 0$, (4.9) gives
\begin{eqnarray*}
S(\lambda'-\lambda;\tau) & \sim &
{\rho \over 2 \tau^{1/2}}
\int_0^\infty du_1 \, e^{- u_1^2 /
 2\gamma} \cos \pi u_1 (\lambda - \lambda')/\tau^{1/2}\\
& & + {\pi \rho (\lambda - \lambda') \over 2 \tau}
\int_0^\infty du_1 \, u_1 e^{- u_1^2 /
 2\gamma} \sin \pi u_1 (\lambda - \lambda')/\tau^{1/2}.
\end{eqnarray*}
Thus, from (2.17)
\renewcommand{\theequation}{4.11}
 \begin{eqnarray}
f(v) & = & {1 \over 2} \int_0^\infty du_1 \,
 e^{- u_1^2 / 2\gamma} \cos \pi u_1 v 
+ {v \over 2}  \int_0^\infty du_1 \, u_1 e^{- u_1^2 / 2\gamma} 
\sin \pi u_1 v \nonumber \\
& = & {1 \over 2} \Big ( {\gamma \over 2 \pi} \Big )^{1/2} e^{- v^2 \gamma / 2}
(1 + \gamma v^2),
\end{eqnarray}
which is the arithmetic means of the distributions (2.18) and (2.20).
\section{Equispaced initial conditions}
In this  section the equal parameter distribution function (2.1) will be
calculated for the solution of the Dyson Brownian motion model for the circular
ensemble with $\beta = 2$ and the initial conditions
$$
\lambda_j^{(0)} = -L/2 + (j-1)/ \rho + \nu, \qquad 0 < \nu < 1/\rho,
\eqno (5.1)
$$
which says the eigenvalues are all equispaced (harmonic
oscillator type spectrum) at $\tau = 0$.

By integrating the Green's function (3.5)  over the initial
configuration (5.1)  (which as a distribution consists of
equispaced delta function peaks), we obtain for the corresponding
eigenvalue p.d.f.
$$
p(\lambda_1, \dots, \lambda_N; \nu,\tau)
= N^{-N/2} e^{E_0 \tau / \gamma} \det [{1 \over L}\theta_1
(\pi ( \lambda_j - (k-1)/\rho - \nu)/L; q)]_{j,k = 1, \dots N}
$$
$$
\times \prod_{1 \le j < k \le N}  \sin \pi (\lambda_k - \lambda_j)/L
\eqno (5.2)
$$

\subsection{Analogy with a classical log-potential system}
Some insight into the expected properties of the two-point
equal parameter distribution for (5.2)  can be obtained from
a known identity [14] expressing the determinant in (5.2), averaged over 
$\nu$, in terms of a product:
\begin{eqnarray*}\lefteqn{
\rho \int_0^{1 / \rho} d \nu \,
\det [{1 \over L} \theta_1 (\pi (\lambda_j/L  -( k - 1) /N + \nu/L );q^{1/N})
]_{j,k = 1,\dots,N}} \\
& & = \left ( {1 \over L } \right )^N f_N(q) \prod_{1 \le j < k \le N}
\theta_1 \left ( {\pi \over L}(\lambda_k - \lambda_j); q \right )
\end{eqnarray*}
where
$$
f_N(q) = N^{N/2} q^{-(N-1)(N-2)/24} \left ( \prod_{k=1}^\infty
(1 - q^{2k}) \right )^{-(N-1)(N-2)/2}
$$
 Thus $p_s$ is proportional to the Jastrow-type
product
$$
 \prod_{1 \le j < k \le N}  \sin { \pi \over L} (\lambda_k - \lambda_j)
\theta_1 \left ( {\pi \over L}(x_k - x_j); q \right )
$$
which can be interpreted as the Boltzmann factor of a classical gas with
potential energy
$$
-  \sum_{1 \le j < k \le N} \log \Big |
 \sin {\pi \over L} (\lambda_k - \lambda_j)
\theta_1 \left ( {\pi \over L}(\lambda_k - \lambda_j); q \right ) \Big |
$$
and $\beta = 1$.

It is well known (see e.g.~[14]) that the $\theta_1$-function is the solution
of the two-dimensional Poisson equation in doubly periodic boundary conditions,
with period $L$ in the $x$-direction and period $W$ in the $y$-direction,
and $q^N = e^{- \pi W /L}$. Furthermore, in the limit $L \rightarrow
\infty$ with $W$ fixed,
$$
-\log \,\theta_1(\pi \lambda /L;q^N) \: \sim \: V_1(\lambda) \qquad {\rm where}
\quad V_1(\lambda) := - \pi |\lambda| /W
$$
while
$$
- \log \,\sin \pi \lambda/L  \: \sim \: V_2(\lambda) \qquad {\rm where}
\quad V_2(\lambda) := - \log |\lambda|
$$
According to Ornstein-Zernicke direct correlation function theory, the
small-$k$ behaviour of the dimensionless structure factor
$$
s(k) := 1 + {1 \over \rho} \int_{-\infty}^\infty dx \,
\rho_{(2)}^T(x) e^{i k x}
\eqno (5.3)
$$
of a classical gas is determined by the small-$k$ behaviour of the reciprocal
of the Fourier transform of the potential:
$$
s(k) \: \sim \: { 1\over \beta \tilde{V}(k)} \qquad {\rm as} \quad 
k \rightarrow 0.
$$  
Now $\hat{V}_1(k) \sim \pi /|k|$ while $\hat{V}_2(k) \sim O(k^2)$.
Hence 
$$
s(k) \: \sim \: |k| / \pi
\eqno (5.4)
$$
or equivalently, for $\lambda \rightarrow \infty$ and $\tau$-fixed
$$
\rho_{(2)}^T(\lambda) \: \sim \: - {1 \over (\pi \lambda)^2}.
$$

\subsection{Exact expression for the equal parameter distribution functions}
To calculate the equal parameter distributions (2.1), we multiply (5.2) by unity in the
form of
$$
i^{-N(N-1)/2} N^{N/2} \det [ e^{2 \pi i j k /N}]_{j = 1, \dots, N \atop
k = -N/2, \dots, N/2 - 1}
$$
to obtain [14],
\renewcommand{\theequation}{5.5}
\begin{eqnarray}\lefteqn{ i^{-N(N-1)/2} \prod_{1 \le j < k \le N}
2 \sin \pi (\lambda_k - \lambda_j)/L} \nonumber \\& & 
\det [{1 \over L} e^{2 \pi i (k + 1/2)(\lambda_j - \nu)/L}
\theta_3( \pi \rho (\lambda_j - \nu) + i \pi^2 \tau N (k + 1/2)/ L^2 \gamma;
q^{N^2})]_{j = 1, \dots,N \atop k = -N/2, \dots,N/2-1}
\end{eqnarray}
The advantage of this expression is that with the product expressed as
a Vandermonde determinant using (3.14), it consists of a product of determinants
in which the elements have an orthogonality property:
$$\int_{-L/2}^{L/2} d\lambda_j \,e^{-2 \pi i (k + 1/2)(\lambda_j - \nu)/L}
e^{2 \pi i (k' + 1/2)(\lambda_j - \nu)/L}
\theta_3( \pi \rho (\lambda_j - \nu) + i \pi^2 \tau N (k' + 1/2)/ L^2 \gamma;
q^{N^2}) = L \delta_{k,k'}
\eqno (5.6 )
$$
This feature allows the distributions (2.1) to be expressed as the determinant
$$
\rho_{(n)}(\lambda_1, \dots, \lambda_N; \nu,\tau) 
= \det [K(\lambda_j, \lambda_k;\nu,\tau)]_{j,k = 1, \dots,N}
\eqno (5.7a)
$$
where
$$
K(\lambda_j, \lambda_k;\nu,\tau) :=\sum_{p = -N/2}^{N/2 - 1}
{1 \over L} e^{2 \pi i (p + 1/2)(\lambda_j - \lambda_k/L)}
\theta_3( \pi \rho (\lambda_k - \nu) + i \pi^2 \tau N (p + 1/2)/ L^2 \gamma; 
e^{-2 \pi^2 \rho^2 \tau / \gamma})
\eqno (5.7b)
$$
The thermodynamic limit follows by noting that this expression is a Riemann
sum approximation to a definite integral, which becomes exact in the
thermodynamic limit, thus giving
$$
K(\lambda_j, \lambda_k;\nu,\tau) = \rho \int_{-1/2}^{1/2}
ds \, e^{- 2 \pi i \rho (\lambda_j - \lambda_k) s} 
 \theta_3( \pi \rho (\lambda_k - \nu) + i \pi^2 \tau
 \rho^2 s/ \gamma; 
e^{-2 \pi^2 \rho^2 \tau / \gamma})
\eqno (5.8)
$$

From (5.7a) and (5.8), the truncated two-point distribution averaged over $\nu$ is
given by
$$
\rho_{(2)}^T(\lambda-\lambda';\tau) = -\rho^2 \int_{-1/2}^{1/2} ds
\int_{-1/2}^{1/2} ds' \, e^{-2 \pi i (\lambda - \lambda')\rho (s - s')}
\hspace{3cm}
$$
$$
\times \theta_3( \pi \rho (\lambda - \lambda') + i \pi^2 \tau
 \rho^2 (s - s')/ \gamma; 
e^{-2 \pi^2 \rho^2 \tau / \gamma})
\eqno (5.9)
$$
From this expression, it is straightforward to compute the small-$|k|$
behaviour of the structure factor (5.3) (only the $n = 0, \pm 1$ terms in
the series definition of $\theta_3$ contribute). We find
$$
s(k) \: \sim \: {|k| \over 2 \pi}(1 + e^{- 2 \pi^2 \tau \rho |k| / \gamma} )\
\eqno (5.10)
$$
For fixed-$\tau$ the term in brackets is to leading order equal to 2, so the
result (5.4) is verified.

\subsection{Asymptotics of the spacing distribution}
The distribution $p(s)$ of the nearest neighbour spacing averaged over $\nu$
can be computed by differentiation of the probability $h(s)$ that an interval
of length $s$ averaged over $\nu$ is free from eigenvalues:
$$
p(s) = -{d^2 \over d s^2} h(s).
$$
The leading asymptotics of $h(s)$ can be predicted by adapting an electrostatic
argument of Dyson [15].

In Section 5.1 we have noted that the eigenvalue p.d.f~(5.2) averaged over
$\nu$ can be interpreted as the Boltzmann factor of a classical gas with
potential which to leading order in large-$\lambda$ behaves as
$$
V_1(\lambda) := - \pi |\lambda|/W, \qquad W = 2 \pi \tau \rho/\gamma.
\eqno (5.11)
$$
Now (5.11) is proportional to the potential of the Coulomb repulsion between
sheets of charge (i.e.~the one-dimensional Coulomb potential), and the
corresponding classical gas is the one-component one-dimensional plasma.
The basis of the electrostatic argument is that in the macroscopic hole
size limit $s \rightarrow \infty$, the plasma behaves like a perfect
conductor. Thus the total charge $- \rho s$ inside the hole due to the
uniform background will be exactly cancelled by two delta function charge
distributions at each boundary of the hole of charge $\rho s/2$.

According to the hypothesis of Dyson 
$$
h(s) \: \sim \: e^{-( U_1(s) + U_2(s) + U_3(s))}
\eqno (5.11)
$$
where $U_1(s), U_2(s)$ and $U_3(s)$ are given by
$$
U_1(s) := -\rho^2 {\pi \over 2W} \int_0^s dx \int_0^s dy \, |x - y|
\eqno (5.12a)
$$
$$
U_2(s) := \rho^2 {s \pi \over 2 W} \Big ( \int_0^s dx \, |x|
+ \int_0^s dy \, |s - y| \Big )
\eqno (5.12b)
$$
$$
U_3(s) := - \Big ( { \rho s \over 2 } \Big )^2 { \pi s\over W}
\eqno (5.12c)
$$
and represent the electrostatic energy of the background-background,
background-boundary and boundary-boundary interactions respectively.
Evaluating (5.12a) and (5.12b) and substituting in (5.11) gives
$$
h(s) \: \sim \: e^{- (\rho s)^3 \gamma / 24 \rho^2 \tau}.
\eqno (5.13)
$$
This is to  be contrasted to the asymptotics of $h(s)$ with $\tau \rightarrow
\infty$  (see e.g.~[1]):
$$
h(s) \: \sim \: e^{- (\pi \rho s)^2 /8}. 
$$
\section{Multiple parameters}
Beenakker and Rejaei [7] considered the multiple-parameter Fokker-Planck equation
$$
{1 \over d} \sum_{\mu = 1}^d \gamma_\mu {\partial P \over \partial \tau_\nu}
={\cal L} P
\eqno (6.1)
$$
with ${ \cal L}$ given by (1.1) and $W$ by (1.2), as a model for the statistical
description of the dispersion relation of a $d$-dimensional crystalline
lattice. With the initial condition (2.7) extended to $\vec{\tau} = \vec{0}$,
it is easy to see that the Green's function solution of (6.1) with $\beta = 2$,
$N$ odd
and $W$ given by (3.1) is obtained by making the replacement
$$
{ 1 \over L} \theta_3(\pi (\lambda_j - \lambda_k)/L;q)
\mapsto {1 \over d} \sum_{\mu = 1}^d 
  \theta_3(\pi (\lambda_j - \lambda_k)/L;q_\mu), \qquad q_\mu := 
e^{- 2 \pi^2 \tau_\mu  /L^2 \gamma_\mu}
\eqno (6.2)
$$

Furthermore, explicit calculation shows that
the density-density correlations and equal-time correlations for the
multiple-parameter equation are the same as for the single-parameter equation
(with corresponding initial eigenvalue p.d.f.'s) provided the replacements
$$
e^{\pm \pi^2 \rho^2 u^2 \tau / 2 \gamma} \mapsto
\left ( {1 \over d} \sum_{\mu = 1}^d 
e^{- \pi^2 \rho^2 u^2 \tau_\mu  / 2 \gamma_\mu} \right )^{\mp 1}
\eqno (6.3)
$$
are made in each of the integrands of (2.11), (3.32) and (4.9), while in (5.7)
the replacement
$$
\theta_3( \pi \rho (\lambda_k - \nu) + i \pi^2 \tau s \rho^2 /  \gamma; 
e^{-2 \pi^2 \rho^2 \tau / \gamma})
\mapsto {1 \over d} \sum_{\mu = 1}^d 
\theta_3( \pi \rho (\lambda_k - \nu) + i \pi^2 \tau_\mu s \rho^2/ \gamma_\mu; 
e^{-2 \pi^2 \rho^2 \tau_\mu / \gamma_\mu})
\eqno (6.4)
$$
is required.

As an example, the multiple parameter generalization of the density-density
correlation (2.11) for initial conditions with unitary symmetry is
$$
S(\lambda - \lambda'; \vec{\tau}) = \rho^2 \int_1^\infty du_1 \,
 \left ( {1 \over d} \sum_{\mu = 1}^d 
e^{- \pi^2 \rho^2 u^2_1 \tau_\mu  / 2 \gamma_\mu} \right )
\cos \pi u_1 \rho (\lambda - \lambda')
$$
$$
\times \int_0^1 du_2 \,
 \left ( {1 \over d} \sum_{\mu = 1}^d 
e^{- \pi^2 \rho^2 u^2_2 \tau_\mu  / 2 \gamma_\mu} \right )^{-1}
\cos \pi u_2 \rho (\lambda - \lambda')
\eqno (6.5)
$$
For large-$|\lambda - \lambda'|$ and $\tau_\mu$ $(\mu = 1, \dots, d)$
(with each $\tau_\mu$ of the same order), the asymptotic expansion of (6.5)
is obtained by writing
$$
 \sum_{\mu = 1}^d 
e^{- \pi^2 \rho^2 u^2_2 \tau_\mu  / 2 \gamma_\mu} =
\exp \Big (- {1 \over d} \sum_{\mu = 1}^d {\tau_\mu \over 2 \gamma_\mu}
(\pi \rho u_2)^2 \Big ) \hspace{3cm}
$$
$$
\times \sum_{\mu = 1}^d \exp \Big (- ({d-1 \over d}{\tau_\mu \over 2 \gamma_\mu}
- {1 \over d} \sum_{\sigma = 1 \atop \sigma \ne \mu}^d
{\tau_\sigma \over 2 \gamma_\sigma})(\pi \rho u_2)^2 \Big )
$$

By the assumption  on the relative order of each $\tau_\mu$, we
see the second term on the r.h.s.~is $O(1)$ and can so be ignored for 
purposes of computing the leading order asymptotic behaviour. Thus we
again find the formula (2.10)  with $\beta = 2$, provided the replacement
$$
{\tau \over 2 \gamma} \: \mapsto \: {1 \over d}  
\sum_{\mu = 1}^d {\tau_\mu \over 2 \gamma_\mu}
\eqno (6.6)
$$
is made. This same prescription gives the asymptotic expansion of the
density-density correlation with multiple parameters for orthogonal and
symplectic symmetry from the corresponding expansion for the single
parameter case obtained in the text. These finding are in agreement with
those in [7] (however the requirement that the expansions are
only valid for each $\tau_\mu$ of the same order was not made explicit in
[7]).

Similarly, we find the distributions (2.17), (2.19), (3.33) and (4.11) are
unchanged, provided the replacement
$$
  \Big ( {\gamma \over 2 \pi} \Big )^{1/2} e^{- v^2 \gamma / 2}
\mapsto {1 \over d} \sum_{\mu = 1}^d  
\Big ( {\gamma_\mu \over 2 \pi} \Big )^{1/2} e^{- v^2 \gamma_\mu / 2}
\eqno (6.7)
$$
is made.

\vspace{1cm}
\noindent
{\bf Note added:} Since the completion of this work, I have been informed by
J.A. Zuk that the results of ref. [3] include the calculation of the
density-density correlations given in Sections 3 and 4. However these results,
which are derived using the supersymmetry method, are presented as complicated
triple integrals, whereas our results (3.30) and (4.9) are products of
simple single integrals. We have not been able to show directly that the
different forms are equivalent. Instead Zuk [16] is repeating the calculation
of [3], with a different parametrization [17] which is known to reproduce
the equal parameter results (2.4)-(2.6), in a bid to derive (3.30) and
(4.9) using the supersymmetry method. He also points out that taking
the Laplace transform of (3.32) in the variable $t:= (\pi \rho)^2\tau/2\gamma$
gives a result entirely in terms of elementary functions:
\begin{eqnarray*}
\tilde{S}(k,s) &:= &\int_{-\infty}^\infty dx \, e^{ikx} \int_0^\infty dt \,
e^{-st}\rho^{-2} S(\lambda ; \tau), \qquad (x := \pi \rho \lambda) \\
&=&{\pi \over 2 |k| (s - k^2)} \left \{
\begin{array}{l}s \log {(1+|k|)^2 + s - 1 \over
-(1 - |k|)^2 + s +1} - k^2 \log(1+|k|), \quad |k| \le 2 \\
s \log {(|k|+1)^2 + s - 1 \over
(|k|-1)^2 + s -1} - k^2 \log{|k|+1 \over |k|-1}, \quad |k| \ge 2
\end{array} \right .
\end{eqnarray*}

We also note that the paper [18] has appeared which states a result
(eq. (47)) equivalent to our (5.9).

\pagebreak

\begin{description}
\item[References]
\item[][1] M.L. Mehta, {\it Random Matrices}, 2nd ed. (Academic, New York,
1991)
\item[][2] Z. Pluha\v{r}, H.A. Weidenm\"uller, J.A. Zuk and C.H. Lewenkopf,
Phys. Rev. Lett. {\bf 73}, 2155 (1994)
\item[][3] N. Taniguchi, A. Hashimoto, B.D. Simons and B.L. Altshuler,
Europhys. Lett. {\bf 27}, 335 (1994)
\item[][4] F. Haake, {\it Quantum Signatures of Chaos} (Springer, Berlin,
1992)
\item[][5] A. Pandey and M.L. Mehta, Comm. Math. Phys. {\bf 87}, 449 (1983);
M.L. Mehta and A. Pandey, J. Phys. A {\bf 16}, 2655 (1983)
\item[][6] A. Pandey and P. Shulka, J. Phys. A {\bf 24}, 3907 (1991)
\item[][7] C.W.J. Beenakker and B. Rejaei, Physica A {\bf 203}, 61 (1994)
\item[][8] A.M.S. Mac\^{e}do, Europhys. Lett. {\bf 27}, 335 (1994)
\item[][9] V.E. Kravstov and M.R. Zirnbauer, Phys. Rev. B {\bf 46},
4332 (1992)
\item[][10] B.D. Simons and B.L. Altshuler, Phys. Rev. B {\bf 48}, 5422 (1993)
\item[][11] F.J. Dyson, J. Math. Phys. {\bf 3}, 1199 (1962)
\item[][12] H. Risken, {\it The Fokker Planck equation}, (Springer, Berlin,
1992)
\item[][13] P.J. Forrester, J. Phys. A {\bf 23}, 1259 (1990)
\item[][14] P.J. Forrester, SIAM J. Math. Anal. {\bf 21}, 270 (1990)
\item[][15] F.J. Dyson, J. Math. Phys. {\bf 3}, 157 (1962)
\item[][16] J.A. Zuk, in preparation
\item[][17] A. Altland, S. Iida and K.B. Efetov, J. Phys. A {\bf 26}, 3545
(1993)
\item[][18] A. Pandey, Chaos, Solitons and Fractals {\bf 5}, 1275 (1995)
\end{description}
\end{document}